\documentstyle[11pt,epsfig,twoside,cite,amssymb]{article}
\newcommand{\bc}{\begin{center}}
\newcommand{\ec}{\end{center}}
\begin{document}
\begin{flushright}
PRA-HEP 00-01
\end{flushright}
\begin{center}
{\Large \bf The relevance of ${\mathbf {\gamma^*_L}}$ in hard
collisions of virtual photons}

\vspace*{0.7cm}
Ji\v{r}\'{\i} Ch\'{y}la and Marek Ta\v{s}evsk\'{y}

\vspace*{0.2cm}
{\em Institute of Physics, Na Slovance 2, Prague 8, Czech Republic}

\vspace*{0.5cm}
{\large \bf Abstract}
\begin{quotation}
\noindent
We explore the relevance of extending the concept of the structure
to the longitudinally polarized virtual photon involved in hard
collisions. We show that for moderate photon virtualities and in the
kinematical region accessible in current experiments at HERA and LEP,
the contributions of its longitudinal polarization to hard collisions
are sizable and should be taken into account as part of the resolved
photon component.
\end{quotation}
\end{center}

\section{Introduction}
In QED quantized in covariant gauge, longitudinally polarized
on--shell photons are present, but due to gauge invariance decouple,
order by order in perturbation theory, in expressions for physical
quantities. For the virtual photon with nonzero virtuality
\footnote{In this paper the virtuality $\tau$ of a particle with
four--momentum $k$
and mass $m$ is defined as $\tau\equiv m^2-k^2$.
In this convention, $P^2>0$ in the space--like region relevant
for hard collisions involving photons in the initial state.}
its longitudinal polarization, denoted $\gamma_L^*$, does,
however, give nonzero contributions to physical
quantities and gauge invariance merely requires that these
contributions vanish as $P^2\rightarrow 0$. In this paper we
discuss hard collisions
\footnote{Characterized by some
``hard scale'' denoted generically as $Q^2$.
In practice ``$Q^2$'' may be standard $Q^2$ in DIS, $E_T^2$
in jet studies, $M_Q^2$ in heavy quark production, etc.}
of virtual photons and, moreover, restrict our attention to
kinematical region $P^2\ll Q^2$ where the concept of virtual photon
structure makes good sense.

This work has been motivated by the lack of the general
attitude toward the role of $\gamma_L^*$ in hard collisions
\footnote{The relevance of $\gamma_L^*$ has recently been
pointed out in \cite{friberg2} as well.}
and in particular by our disagreement with the statements of a
recent paper \cite{GRSch}, where the treatment
of virtuality dependence of physical quantities is based on the
following two claims
\footnote{$f^L_{\gamma(P^2)/e}$ in the notation of \cite{GRSch}
corresponds to $f_L^{\gamma}(P^2)$ in our formula (\ref{fluxL}).}
\begin{description}
\item{(i)}
effects of $f^L_{\gamma(P^2)/e}$ should be neglected since the
corresponding longitudinal cross--sections are suppressed by powers
of $P^2/Q^2$, and
\item{(ii)} cross--sections of partonic subprocesses involving
$\gamma(P^2)$ should be calculated as if $P^2=0$ due (partly) to
the $P^2/Q^2$ suppression of any additional terms.
\end{description}
In the next Section we first analyze these claims and
point out why they are wrong. Then we recall the
reasons for introducing the concept of virtual photon structure
and recollect basic formulae concerning the structure
of $\gamma_L^*$. Numerical results illustrating the importance of
including the contributions of $\gamma_L^*$ are presented in
Section 3.
The feasibility of extracting the information of partonic
content of $\gamma^*_L$ from jet production at HERA is addressed
in Section 4, followed by the summary and conclusions in Section 5.

\section{Theoretical considerations}

\subsection{Virtuality dependence of $\sigma(\gamma^*_L)$}
Before recalling practical usefulness of the concept of partonic
structure of virtual photons, let us show why the claims made in
\cite{GRSch} and mentioned in the Introduction are incorrect.
The fact that in the resolved photon channel the cross--sections
of $\gamma_L^*$ are
not suppressed by $P^2/Q^2$ follows directly from
analysis of the formula (E.1)  in \cite{Budnev} for the cross--section
$\sigma_{TL}$ (denoted $\sigma_{TS}$ there),
which shows that for small $P^2\ll m_q^2$ ($m_q$ being the quark mass)
its contribution to $F_2^{\gamma}(x,P^2,Q^2)$ behaves as
\footnote{In the notation of \cite{Budnev} the first and second
subscripts in $\sigma_{ij}$ refer to polarizations of the
probing and target photons respectively, with wirtualities $Q^2$ and
$P^2$. Most of the terms in the expression for $\sigma_{TL}$ do,
indeed, behave as $P^2/Q^2$, but there is one, proportional to
$(\Delta t q_1^4/T)$, which does not and which yields (\ref{lowP2}).}
\begin{equation}
F^{\gamma}_{TL}(x,P^2,Q^2)=
\frac{P^2}{m_q^2}\frac{\alpha}{\pi}4x^3(1-x)^2
=\frac{P^2}{m^2_q}2x\frac{\alpha}{2\pi}4x^2(1-x)^2.
\label{lowP2}
\end{equation}
This expression coincides, apart from the factor $N_ce_q^2$
appropriate to a quark with electric charge $e_q$ and $N_c$ colors,
with the QED expression for distribution
function of quarks inside $\gamma_L^*$ in our formulae
(\ref{threshold}) below \footnote{In practical
applications the factorization scale $M^2$ in (\ref{finalresult})
is identified with the generic hard scale $Q^2$.} multiplied by
$2x(\alpha/2\pi)$. For $P^2\gg m^2_q$, on the other hand, the
distribution function (\ref{quarklong})
is proportional to $4x(1-x)$ with no $P^2/Q^2$ suppression. As a
result, in the region $P^2\gg m^2_q$, $\gamma_L^*$ supplies finite
contribution to $F_2^{\gamma}(x,P^2,Q^2)$ equal to
$(\alpha/\pi)(N_ce_q^2)4x(1-x)$ \cite{russians}.
Similarly, also the second claim (ii)
is incorrect, because also part of the contribution of
$\sigma_{TT}$ has the same $P^2$ behaviour as in (\ref{lowP2}).
Physical explanation of this behaviour is simple: even for large
values of $Q^2$ the virtuality $\tau$ of the quarks
(antiquarks) from the primary splitting
$\gamma^*\rightarrow q\overline{q}$ of the target photon comes
predominantly from the region
close to its minimal value $\tau^{\mathrm{min}}=xP^2+m_q^2/(1-x)$
and therefore the threshold behaviour is governed by the quark mass
$m_q$ rather than $Q^2$.

On the other hand, virtuality dependence of the contributions of
$\gamma^*_T$ and $\gamma^*_L$ can be safely neglected in the LO
direct photon hard processes, for instance in large $E_T$ jet
production via the photon--gluon fusion subprocess
$\gamma^* G\rightarrow q\overline{q}$. In these processes virtuality
of the exchanged quark (or antiquark) is forced by kinematics
to be proportional to jet transverse energy $E_T$ and therefore the
virtuality dependent part is suppressed by powers of $P^2/E_T^2$.
Of course, in realistic QCD the onset of quark distribution functions
of $\gamma^*_L$ is not expected to be determined directly by quark
masses, but rather by some nonperturbative parameter related to
confinement, but the basic features of the dependence on $P^2$,
exemplified in (\ref{lowP2}), are likely to persist.

\subsection{Equivalent photon approximation}
Most of the present knowledge of the structure of the photon
comes from experiments at the ep and e$^+$e$^-$ colliders, where
the incoming leptons act as sources of transverse and longitudinal
virtual photons. To order $\alpha$ their respective
unintegrated fluxes are given as
\begin{eqnarray}
f^{\gamma}_{T}(y,P^2) & = & \frac{\alpha}{2\pi}
\left(\frac{1+(1-y)^2)}{y}\frac{1}{P^2}-\frac{2m_{\mathrm e}
^2 y}{P^4}\right),
\label{fluxT} \\
f^{\gamma}_{L}(y,P^2) & = & \frac{\alpha}{2\pi}
\frac{2(1-y)}{y}\frac{1}{P^2}.
\label{fluxL}
\end{eqnarray}
Phenomenological analyses of interactions of virtual photons have so
far concentrated on its transverse polarization. The same holds for
available parameterizations of parton distribution functions (PDF) of
virtual photons. Neglecting longitudinal photons is in general a good
approximation for $y\rightarrow 1$, where the flux
$f_L^{\gamma}(y,P^2)\rightarrow 0$, as well as for very small
virtualities $P^2$, where PDF of $\gamma_L^*$ vanish by gauge
invariance. But how small is ``very small'' in fact? For instance,
should we take into account the contribution of $\gamma^*_{L}$ to
jet cross--section in the region $E_T\gtrsim 5$
GeV, $P^2\gtrsim 1$ GeV$^2$, where most of the data on virtual
photons obtained in ep collisions at HERA come from? The rest of this
paper is devoted to addressing this and related questions.

\subsection{Who needs the concept of partonic structure of virtual
photons?}
Let us briefly recall the virtue of extending the concept of
partonic ``structure'' to virtual photons. The arguments for it
were discussed in detail in \cite{smarkem1,smarkem2,smarkem3} and
we therefore merely summarize the most important points:
\begin{itemize}
\item In principle, the concept of partonic structure of (sufficiently)
virtual photons can be dispensed with because higher order
perturbative QCD corrections to cross--sections of processes
involving virtual
photons in the initial state are well--defined and finite even for
massless partons.
\item In practice, however, this concept is extraordinarily
useful as it allows us
to include the resummation of higher order QCD effects
that come from physically well--understood region of (almost)
parallel emission of partons off the quark (or antiquark) coming
from the primary $\gamma^*\rightarrow q\overline{q}$ splitting
and subsequently participating in hard processes.
\end{itemize}
In other words, for the virtual photon, as opposed to the real one,
its PDF can be regarded as ``merely'' describing higher order
perturbative effects and not the ``true'' structure. Although this
distinction between the content of PDF of real and virtual photons
does exist, it does not affect the extraordinary
{\em phenomenological} usefulness of PDF of the virtual photon.
As shown in \cite{smarkem1,smarkem2,smarkem3}
the nontrivial part of the resolved photon contributions to NLO
calculations of dijet production at HERA obtained with JETVIP
\cite{JETVIP} is large and affects significantly the conclusions of
phenomenological analyses of existing experimental data.

\subsection{Structure of $\gamma_L^*$ in QED}
The definition and evaluation of quark distribution functions of
the virtual
photon in QED serves as a guide to QCD improved parton model
predictions of virtuality dependence of their pointlike parts.
In pure QED and to order $\alpha$ the probability of
finding inside $\gamma_T^*$ or $\gamma_L^*$ of virtuality $P^2$ a
quark with mass
$m_q$, electric charge $e_q$, momentum fraction $x$ and virtuality
$\tau\le M^2$, is given, in units of $3e_q^2\alpha/2\pi$,
as \cite{smarkem3} ($k=T,L$)
\begin{equation}
q^{\mathrm {QED}}_k(x,m_q^2,P^2,M^2)=
f_k(x)\ln\left(\frac{M^2}{\tau^{\mathrm {min}}}\right)+
\left[-f_k(x)+\frac{g_k(x)m_q^2+h_k(x)P^2}{\tau^{\mathrm {min}}}
\right]
\left(1-\frac{\tau^{\mathrm {min}}}{M^2}\right),
\label{fullresult}
\end{equation}
where $\tau^{\mathrm {min}}=xP^2+m_q^2/(1-x)$.
The quantity defined in (\ref{fullresult}) has a clear physical
interpretation: it describes the flux
of quarks and antiquarks that are almost collinear with the incoming
photon and ``live'' longer
\footnote{In fact most of these quarks live much longer than $1/M$.}
than $1/M$. For $\tau^{\mathrm {min}}\ll
M^2$ the expression (\ref{fullresult}) simplifies to
\begin{equation}
q^{\mathrm {QED}}_{k}(x,m_q^2,P^2,M^2)  =  f_k(x)\ln\left(
\frac{M^2}{xP^2+m_q^2/(1-x)}\right)-f_k(x)+
\frac{g_k(x)m_q^2+h_k(x)P^2}{xP^2+m_q^2/(1-x)},
\label{finalresult}
\end{equation}
which for $x(1-x)P^2\gg m_q^2$ reduces further to
\begin{equation}
q^{\mathrm {QED}}_k(x,0,P^2,M^2)=f_k(x)\ln\left(\frac{M^2}{xP^2}\right)
-f_k(x)+\frac{h_k(x)}{x}.
\label{virtualphoton}
\end{equation}
The functions $f_k,g_k,h_k$ are given as \cite{smarkem3}
\begin{equation}
\begin{array}{lll}
 f_T(x)=x^2+(1-x)^2, & g_T(x)=
{\displaystyle
 \frac{\displaystyle 1}{\displaystyle 1-x}
}, & h_T(x)=0, \\[0.4cm]
 f_L(x)=0, & g_L(x)=0, & h_L(x)=4x^2(1-x).
\label{fghTL}
\end{array}
\end{equation}
For $M^2\gg x(1-x)P^2$ the quark distribution function of
$\gamma_L^*$ has a simple form
\begin{eqnarray}
q_L^{\mathrm {QED}}(x,m_q^2,P^2,M^2)=
\frac{\displaystyle 4x^2(1-x)^2P^2}{\displaystyle x(1-x)P^2+m_q^2}&
\rightarrow &
4x(1-x);~~~~~~~~~~~~x(1-x)P^2\gg m_q^2\label{quarklong}\\[0.2cm]
& \rightarrow & \frac{\displaystyle P^2}{m_q^2}4x^2(1-x)^2;
~~~~~x(1-x)P^2\ll m_q^2
\label{threshold}
\end{eqnarray}

\subsection{QCD corrections}
For $\gamma_T^*$ QCD corrections to
QED formula (\ref{virtualphoton}) are well understood.
Though important, in particular for large and very small $x$,
they do not change its basic features and the main nontrivial
effect comes from the emergence of gluons inside $\gamma_T^*$.
For $\gamma_L^*$ the effects of collinear parton radiation off
the quarks/antiquarks from the $\gamma^*_L\rightarrow q\overline{q}$
splitting result in factorization scale dependence that resembles
those of hadrons and will be discussed in separate paper. For the
purpose of this exploratory study we use the QED formula
(\ref{quarklong}) throughout this paper.

\section{Numerical results}

\subsection{DIS on $\gamma^*$ in QED}
\begin{figure}\centering
\epsfig{file=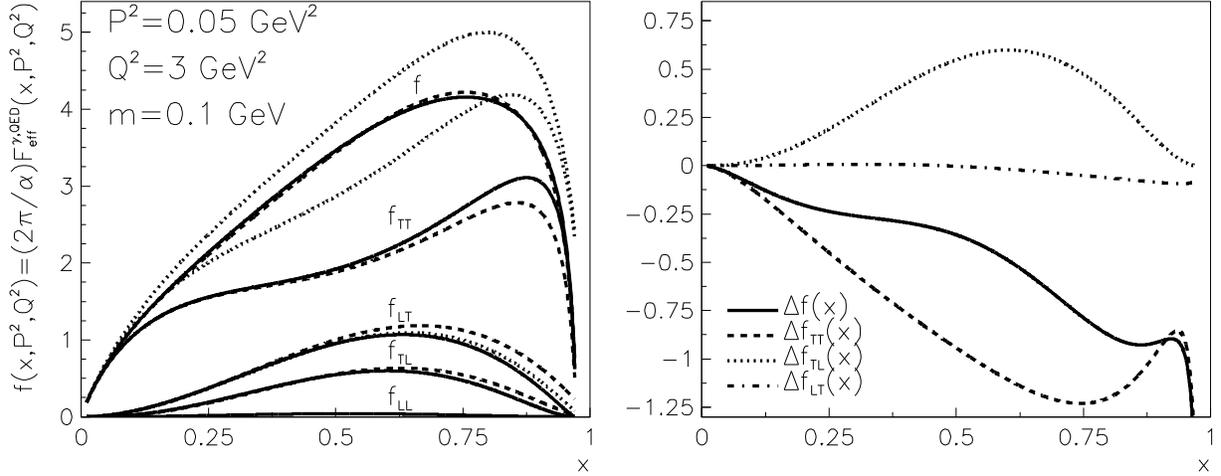,width=\textwidth} \caption{a)
$(2\pi/\alpha)F_{\mathrm{eff}}^{\gamma,\mathrm{QED}}(x,P^2,Q^2)$
evaluated from
eq. (\ref{Feff}) with $\sigma_{jk}$ given by exact QED formulae
(E.1) of \cite{Budnev} (upper solid curve), together with
contributions of individual channels $\sigma_{jk}$ (other solid
curves). The approximate expressions based on the formula
(\ref{finalresult}) as well as the exact ones corresponding to
$P^2=0$ are shown as dashed and dotted curves, respectively. b)
The corresponding differences $\Delta f(x,P^2,Q^2)$ and $\Delta
f_{jk}(x,P^2,Q^2)$.}
\label{OPALfig}
\end{figure}

The cleanest evidence of the importance of taking into account the
contribution of $\gamma_L^*$ has been provided by the L3 and OPAL
measurements \cite{L3,OPAL} of the QED structure function
$F_2^{\gamma,\mathrm{QED}}$ at LEP.
In these measurements, based on the analysis of $\mu^+\mu^-$ final
states, the average target photon virtuality is small
($\langle P^2\rangle=0.033$ GeV$^2$ in \cite{L3} and
$\langle P^2\rangle=0.05$ GeV$^2$ in \cite{OPAL})
but still sufficiently large with respect to $m_{\mu}^2\doteq 0.01$
GeV$^2$ to see the decrease of
$F_2^{\gamma,\mathrm{QED}}(x,P^2,Q^2)$ with respect to the QED
prediction for the real photon. To order $\alpha$ these predictions
were calculated exactly in \cite{Budnev} and contain contributions
of both transverse and longitudinal polarizations of the target
photon. In the region $m_e^2\ll P^2\ll Q^2$ experiments at LEP
actually measure the following sum of $\gamma^*\gamma^*$
cross--sections, the first and second indices corresponding to probe
and target photon respectively,
\begin{equation}
F_{\mathrm{eff}}^{\gamma}(x,P^2,Q^2)\equiv \frac{Q^2}{4\pi^2\alpha}
\left(\sigma_{TT}+\sigma_{LT}+
\sigma_{TL}+\sigma_{LL}\right)
=\frac{Q^2}{4\pi^2\alpha}\sigma(P^2,Q^2,W^2),
\label{Feff}
\end{equation}
where all cross--sections $\sigma_{jk}$ are functions of $W^2,P^2$
and $Q^2$ and $x=Q^2/(W^2+Q^2+P^2)$. As shown in \cite{L3,OPAL} the
data are in very good agreement with QED prediction for (\ref{Feff})
provided the dependence on target photon virtuality $P^2$ is taken
into account. For OPAL kinematical region the QED predictions for
$f(x,P^2,Q^2)\equiv (2\pi/\alpha)F^{\gamma}_{\mathrm{eff}}(x,P^2,Q^2)$
as well as the individual contributions $f_{ij}$, are shown in
Fig. \ref{OPALfig}a, together with the results (shown as dotted
curves) corresponding to the real photon and the approximations using
formulae of the preceding Section (dashed curves). The variations of
$f_{jk}(x,P^2,Q^2)$ and $f(x,P^2,Q^2)$ with respect to real photon,
defined as
($i,j=T,L$)
\begin{equation}
\Delta f_{jk}(x,P^2,Q^2)\equiv f_{jk}(x,P^2,Q^2)-f_{jk}(x,0,Q^2)),
\label{deltaf}
\end{equation}
are plotted in Fig. \ref{OPALfig}b.
The contribution $\Delta f_{TL}=f_{TL}$ to the
variation $\Delta F_{\mathrm{eff}}^{\gamma}(x,P^2,Q^2)$ coming from
target $\gamma_L^*$ is clearly comparable in magnitude to
$\Delta f_{TT}$ and $\Delta f_{LT}$ coming from target $\gamma_T^*$.
Neglecting $\Delta f_{TL}$ would thus lead to serious disagreement
between QED predictions and data.

\subsection{DIS on $\gamma^*$ in QCD}
In LO QCD the structure function $F_2^{\gamma}$ is given in
terms of quark
distribution functions by the same expression as for hadrons
\footnote{In the present paper we disregard the consequences of
the reformulation of QCD analysis of $F_2^{\gamma}$ proposed
in \cite{jfactor} as they do not concern the main point of our
discussion.}
\begin{equation}
F_2^{\gamma}(x,P^2,Q^2)=  \sum_{i}2x e_i^2
\left(q_i(x,P^2,Q^2)+\overline{q}_i(x,P^2,Q^2)\right).
\label{F2PM}
\end{equation}
\begin{figure}[t]
\epsfig{file=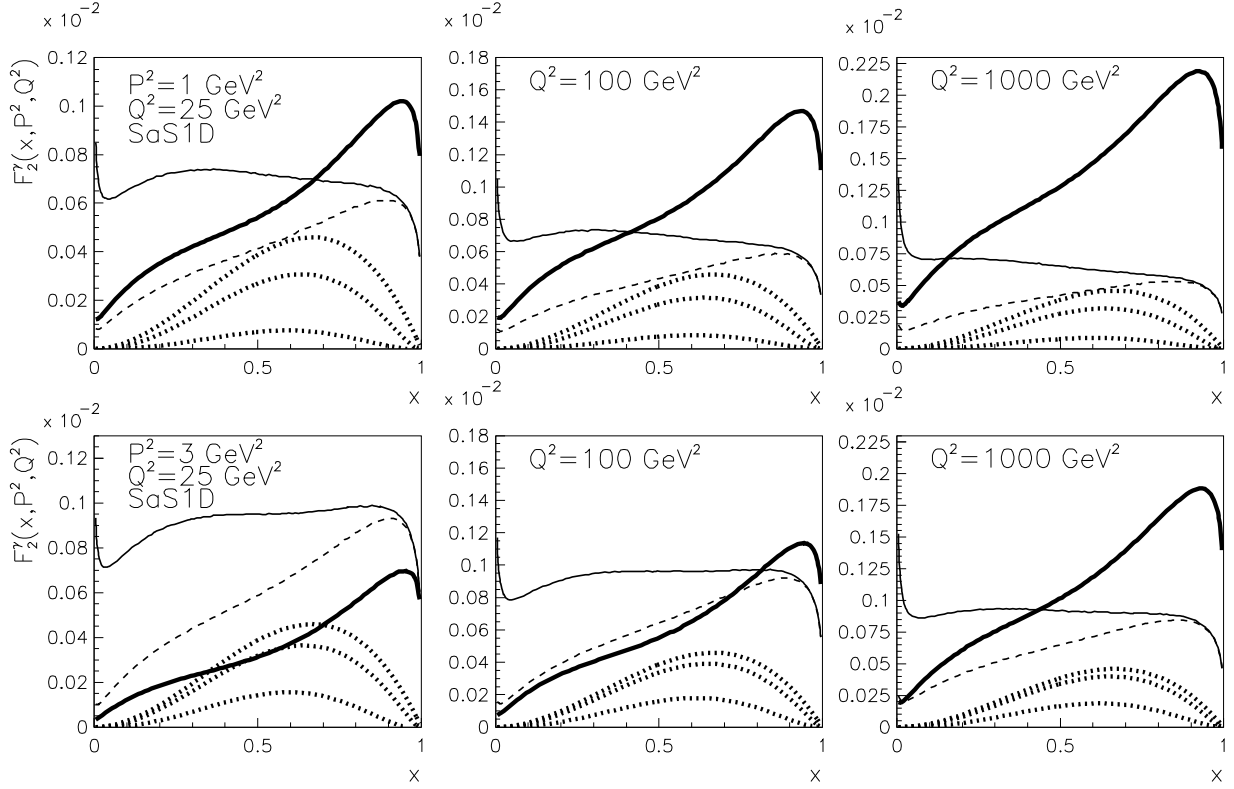,width=\textwidth}
\caption{SaS1D parameterization of $F_2^{\gamma}(x,P^2,Q^2)$
(thick solid curves) compared to QED formula (\ref{quarklong})
for, from above, $m^2=0,0.1$ and $1$ GeV$^2$ (thick dotted curves).
The thin solid curves correspond to
$\Delta F_2^{\gamma_T}(x,P^2,Q^2)$,
the dashed ones to the difference (\ref{DeltaF}) of
pointlike parts of $\gamma_T^*$ only.}
\label{f2efftl}
\end{figure}
In all existing phenomenological analyses of experimental data only
target $\gamma^*_T$ has been taken into account and to the best of
our knowledge no attempt has been made to extract PDF of
$\gamma_L^*$ therefrom. In this exploratory study we compare the
results for $F_2^{\gamma}$ obtained with Schuler--Sj\"{o}strand
(SaS) parameterization \cite{sas1} of $q_T(x,P^2,M^2)$ with the QED
prediction (\ref{quarklong}) for $q_L(x,P^2,M^2)$.

In Fig. \ref{f2efftl} this comparison is performed for typical
values of $P^2$ and $Q^2$ accessible at LEP and $m_q^2=1,0.1$
GeV$^2$ and $m_q^2=0$.
The importance of the contributions of $\gamma_L^*$ with respect to
those of $\gamma_T^*$ depends sensitively on the value of $m_q$:
whereas for $m_q\doteq 1$ GeV, $\gamma_L^*$ is largely irrelevant,
for $m_q\lesssim 0.3$ GeV, medium values of $x$ and $Q^2\lesssim 100$
GeV$^2$, its contributions in the considered region of $P^2$ and
$Q^2$ are comparable to those of SaS1D parameterization of $\gamma_T^*$.
Only for very large $Q^2$ does $\gamma_L^*$ become really negligible
with respect to $\gamma_T^*$. For fixed $Q^2$ the relative importance
of $\gamma_L^*$ with respect to $\gamma_T^*$ grows with $P^2$, but
to retain clear physical meaning of PDF we stay throughout this paper
in the region where $P^2\ll Q^2$.

The comparison of the contributions of $\gamma_L^*$ and
$\gamma_T^*$ at the same values of $P^2$ is one measure of
the relevance of $\gamma_L^*$. If we are interested in
virtuality dependence of $F_2^{\gamma}(x,P^2,Q^2)$, the
appropriate comparison is with difference
\begin{equation}
\Delta F_2^{\gamma_T}(x,P^2,Q^2)\equiv F_2^{\gamma_T}(x,0,Q^2)-
  F_2^{\gamma_T}(x,P^2,Q^2),
\label{DeltaF}
\end{equation}
of SaS results for $\gamma_T^*$, denoted in Fig. \ref{f2efftl} by
thin solid curves. At small to moderate $x$, lower hard scales
$Q^2$ and larger virtualities $P^2$, the contributions of
$\gamma_L^*$ appear by this measure less important
than when compared to $F_2^{\gamma}(x,P^2,Q^2)$ itself. However,
this is due largely to the fact that $F_2^{\gamma}$ of the real
photon gets a large contribution from its VDM component, whereas the
parameterization of $q_L$ used in this comparison corresponds to
purely pointlike expression (\ref{quarklong}). Compared to the
difference (\ref{DeltaF}) of the pointlike parts of $\gamma_T^*$ only,
denoted by dashed curves in Fig. \ref{f2efftl}, the contributions
of $\gamma_L^*$ are, at least for $m_q\lesssim 0.3$ GeV, again quite
significant throughout large part of the kinematical range considered.

\subsection{LO calculations of dijet production in ep and e$^+$e$^-$
collisions}
\begin{figure}[t]
\epsfig{file=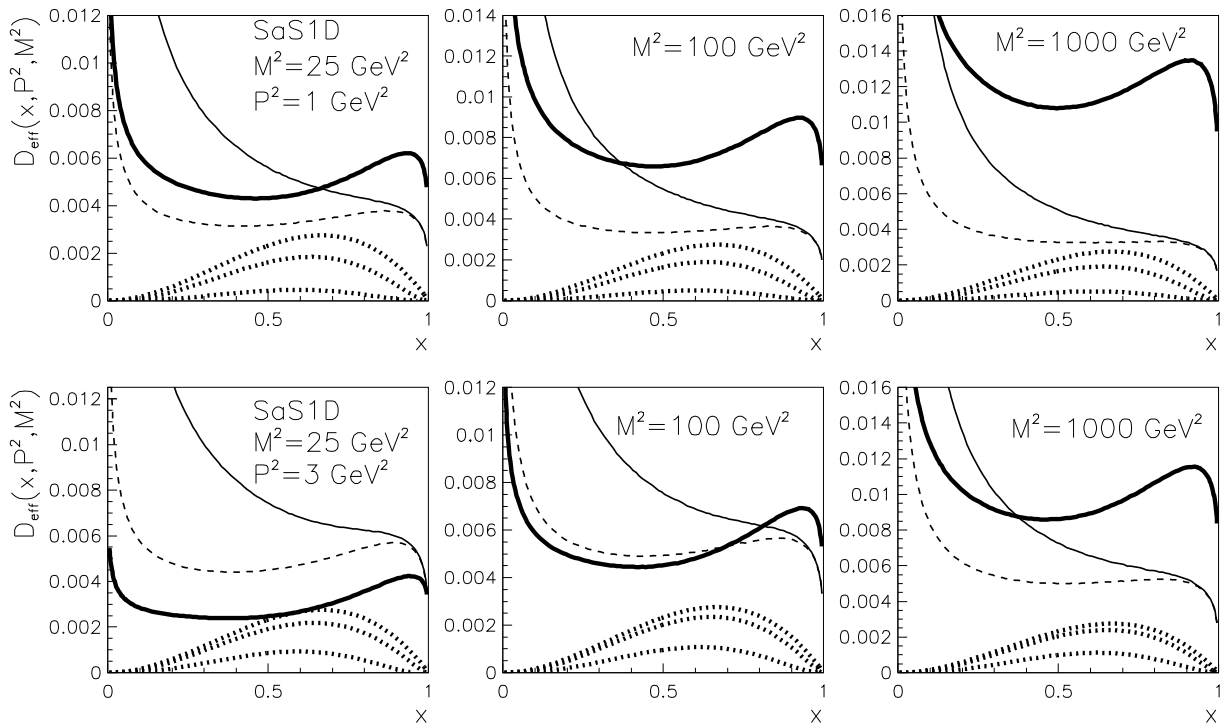,width=\textwidth}
\caption{The same as in Fig. \ref{f2efftl} but for the quantity
$D_{\mathrm eff}(x,P^2,M^2)$ defined in (\ref{deff}).}
\label{f2defftl}
\end{figure}

The measurement of dijet production in ep and e$^+$e$^-$ collisions
provides another way of investigating interactions of the virtual
photon \cite{H1eff,phd}. In general the cross--sections for dijet
production are given as sums of contributions of all possible parton
level subprocess. To demonstrate the importance of including the
contributions of target $\gamma_L^*$ it is, however, sufficient to
use the approximation of the single effective subprocess \cite{Chris}
in which dijet cross--sections are expressed in terms of the so called
{\em effective parton distribution function} of the target photon
\begin{equation}
D_{\mathrm eff}(x,P^2,M^2) \equiv
\sum_{i=1}^{n_f}\left(q_i(x,P^2,M^2)+\overline{q}_i(x,P^2,M^2)\right)+
\frac{9}{4}G(x,P^2,M^2).
\label{deff}
\end{equation}
In Fig. \ref{f2defftl} we perform for this quantity the same
comparisons as we did in Fig. \ref{f2efftl} for $F_2^{\gamma}$,
including the comparison with the difference
$\Delta D_{\mathrm{eff}}(x,P^2,Q^2)$, defined analogously to
(\ref{DeltaF}).
The fact that in QED $\gamma_L^*$ contains
no gluons is reflected in substantially smaller relative
importance of $\gamma_L^*$ for $D_{\mathrm{eff}}$ at small
values of $x$. Otherwise, however, the messages of Figs.
\ref{f2efftl} and \ref{f2defftl} are the same: in hard
processes the relative importance of the contributions of
target $\gamma_L^*$ with respect to those of $\gamma_T^*$
\begin{itemize}
\item depends sensitively on the value of $m_q$,
\item peaks around $x\doteq 0.6$ and vanish for $x\rightarrow 0$ and
$x\rightarrow 1$,
\item grows with target photon virtuality $P^2$ and
\item decreases with factorization scale $M^2$.
\end{itemize}
For physically reasonable value $m_q=0.3$ GeV, Figs. \ref{f2efftl}
and \ref{f2defftl} suggest that at least in part of the kinematical
range accessible at HERA $\gamma_L^*$ should definitely be taken into
account.

\subsection{NLO calculations of dijet production in ep collisions}
\begin{figure}\centering
\epsfig{file=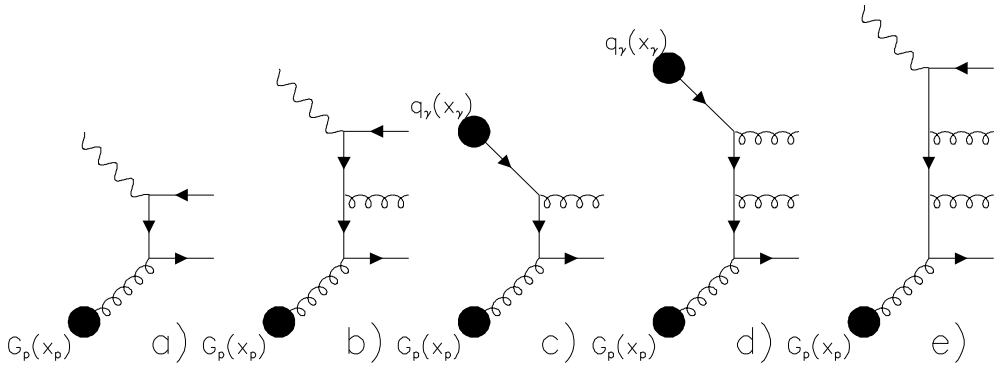, width=14cm}
\caption{Examples of diagrams contributing to dijet production in
ep collisions at the orders $\alpha\alpha_s$ (a), $\alpha\alpha_s^2$
(b,c), and $\alpha\alpha_s^3$ (d,e) taking into account that the upper
blobs representing quark distribution functions of the photon are
proportional to $\alpha$.}
\label{diagrams}
\end{figure}

In the preceding subsections we have discussed the importance of
including the contributions of $\gamma_L^*$ to QED or the LO QCD
quantities $F_{\mathrm{eff}},F_2^{\gamma}$ and $D_{\mathrm{eff}}$.
In this subsection we shall address the same question within
the NLO QCD parton level calculations of dijet cross--sections in ep
collisions, obtained with JETVIP \cite{JETVIP}, currently the only
NLO parton level MC program that includes both direct and resolved
photon contributions
\footnote{In specifying the powers of $\alpha$ corresponding to
various diagrams we discard one common power of $\alpha$
coming from the vertex where the incoming electron emits the virtual
photon. This vertex is also left out in diagrams of
Fig. \ref{diagrams}}.
JETVIP contains the full set of
partonic cross--sections for the direct photon contribution up the
order $\alpha\alpha_s^2$. Examples of such diagrams are in Fig.
\ref{diagrams}a ($\alpha\alpha_s$ tree diagram) and
Fig. \ref{diagrams}b ($\alpha\alpha_s^2$ tree diagram).
To go one order of
$\alpha_s$ higher and perform complete calculation of the direct
photon contributions up to order $\alpha\alpha_s^3$ would require
evaluating tree diagrams like that in Fig. \ref{diagrams}e, as well
as one--loop corrections to diagrams like in Fig. \ref{diagrams}b and
two--loop corrections to diagrams like in Fig. \ref{diagrams}a.
So far, such calculations are not available.
In addition to complete $O(\alpha\alpha_s^2)$ direct photon
contribution JETVIP
includes also the resolved photon one with partonic
cross--sections up to the order
$\alpha_s^3$, exemplified by diagrams in Fig. \ref{diagrams}c,d.
The justification for including in the resolved channel terms of
the order $\alpha_s^3$ are discussed in detail in
\cite{smarkem1,smarkem2,smarkem3}.
Once the concept of virtual photon structure is introduced, part
of the direct photon contribution (which for the virtual photon
is actually nonsingular) is subtracted and included in the definition
of PDF
appearing in the resolved photon contribution. For $\gamma_T^*$ the
subtracted term is given as the convolution of the splitting function
\footnote{JETVIP works with massless quarks and includes in
(\ref{splitterm}) additional function of $x$.}
\begin{equation}
q^{\mathrm split}_T(x,P^2,M^2)=q^{\mathrm{QED}}_T(x,P^2,Q^2)=
\frac{\alpha}{2\pi}3e_q^2\left(x^2+(1-x)^2\right)\ln\frac{M^2}{xP^2}.
\label{splitterm}
\end{equation}
with $\alpha_s^2$ partonic cross--sections.
To avoid misunderstanding we shall henceforth use the term ``direct
unsubtracted'' (DIR$_{\mathrm{uns}}$) to denote NLO direct photon
contributions {\em before} this subtraction and reserve the term
``direct'' for the results {\em after} it. In this terminology the
complete JETVIP calculations are given by the sum of direct and resolved
parts and denoted DIR$+$RES. In JETVIP only the terms defining quark
distribution function of the transverse virtual photon are subtracted
from DIR$_{\mathrm{uns}}$ calculations.

In \cite{smarkem1,smarkem2,smarkem3} we discussed dijet
cross--sections calculated by means of JETVIP in the kinematical
region typical for HERA experiments
$$E_T^{(1)}\ge E_T^c+\Delta,~E_T^{(2)}\ge E_T^c,
~~~E_T^c=5~{\mathrm{GeV}},
~~\Delta=2~{\mathrm{GeV}}$$
 $$-2.5 \le \eta^{(i)}\le 0,~i=1,2,$$
in four windows of photon virtuality
$$1.4\le P^2\le 2.4~{\mathrm {GeV}}^2;
~2.4\le P^2\le 4.4~{\mathrm {GeV}}^2;~
4.4\le P^2\le 10~{\mathrm GeV}^2;
~10\le P^2\le 25~{\mathrm {GeV}}^2$$
and for $0.25\le y\le 0.7$. The whole analysis has been performed in
$\gamma^*$p CMS.
The cuts on $E_T$ were chosen in such a way that
in all $P^2$ windows $\langle P^2\rangle \ll E^2_T$,
thereby ensuring that the virtual photon lives long enough for its
``structure'' to develop before the hard scattering takes place.
The asymmetric cut in $E_T$ is appropriate for our decision to plot
the sums of $E_T$ and $\eta$ distributions of the jets with highest
and second highest $E_T$.
In JETVIP jets are defined by means of the standard cone algorithm
with jet momenta defined using the $E_T$--weighting recombination
procedure and supplemented with the $R_{\mathrm{sep}}$ parameter.
All calculations presented below correspond to $R_{\mathrm{sep}}=2$
and were obtained setting the renormalization scale $\mu$
as well as the factorization scale $M$ equal to jet transverse energy.
The sensitivity to these parameters as well as other ambiguities
are discussed in detail in \cite{smarkem3,phd}.

Beside the splitting term (\ref{splitterm}), which generates quark
distribution function of $\gamma_T^*$, one can subtract from NLO direct
photon calculations also the integral over the term proportional to
$h_L(x)$ and put it into the definition of quark distribution function of
$\gamma_L^*$. To do that properly would, however, require modifying the
original code in order to take into account different $y$ dependence of
the fluxes of $\gamma_T^*$ and $\gamma_L^*$ in (\ref{fluxT}-\ref{fluxL}).
In this exploratory study we neglect this difference and fake the
contributions of $\gamma_L^*$ simply by running JETVIP in the
resolved photon channel using (\ref{quarklong}) with $m_q=0$ as the
input PDF. As in the considered region $\langle y\rangle\doteq 0.4$,
the error incurred by this approximation does not exceed $16$\%.

\begin{figure}\unitlength 1mm
\begin{picture}(160,160)
\put(0,85){\epsfig{file=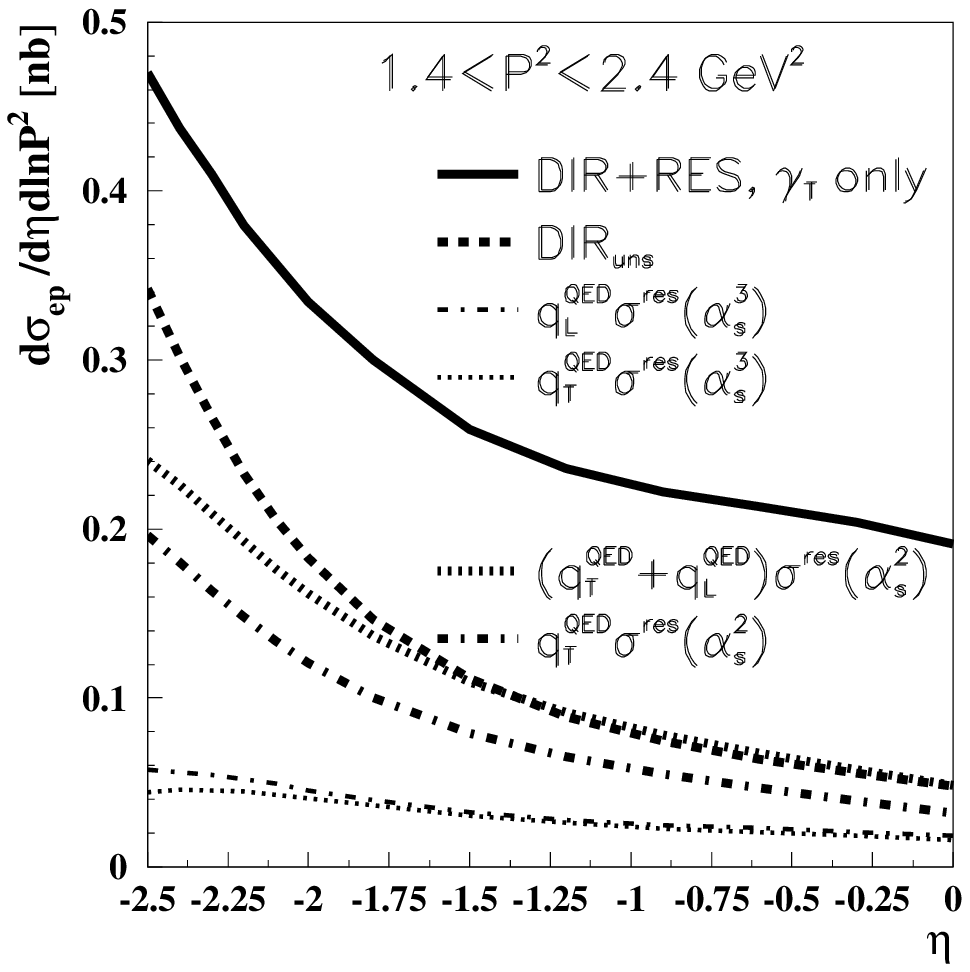,width=8.5cm,height=8.5cm}}
\put(85,85){\epsfig{file=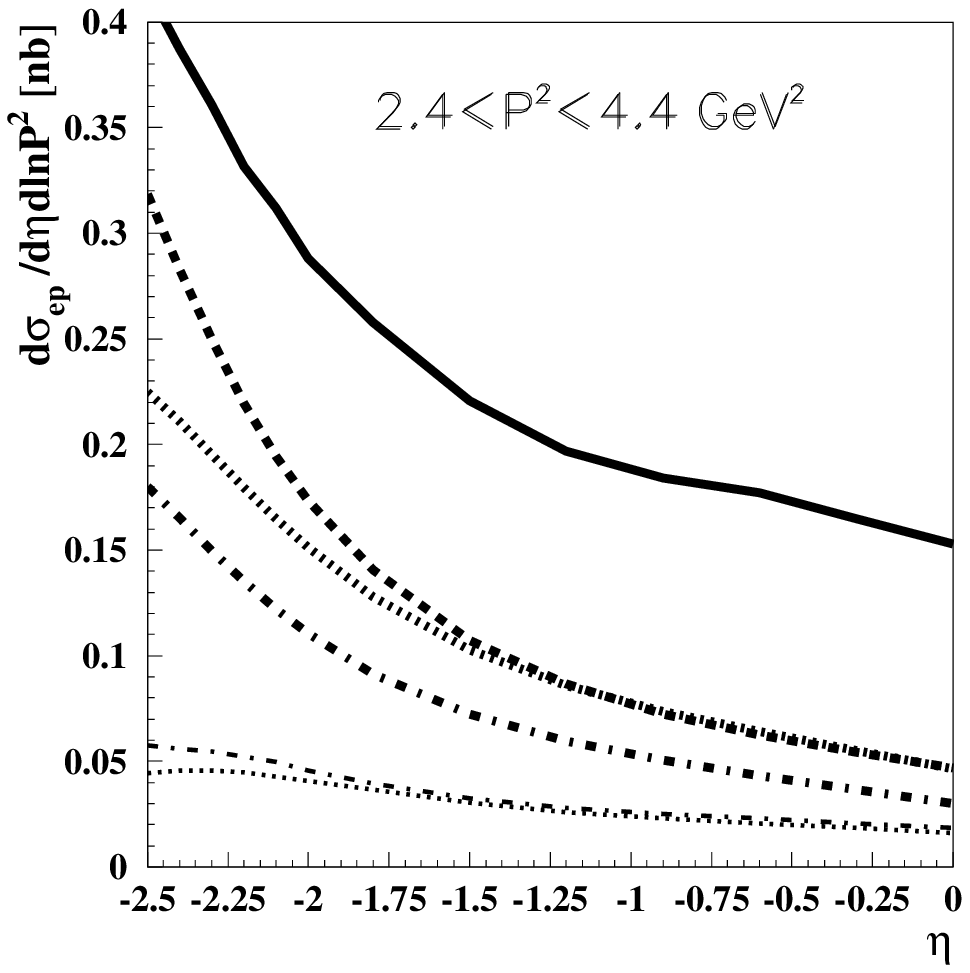,width=8.5cm,height=8.5cm}}
\put(0,0){\epsfig{file=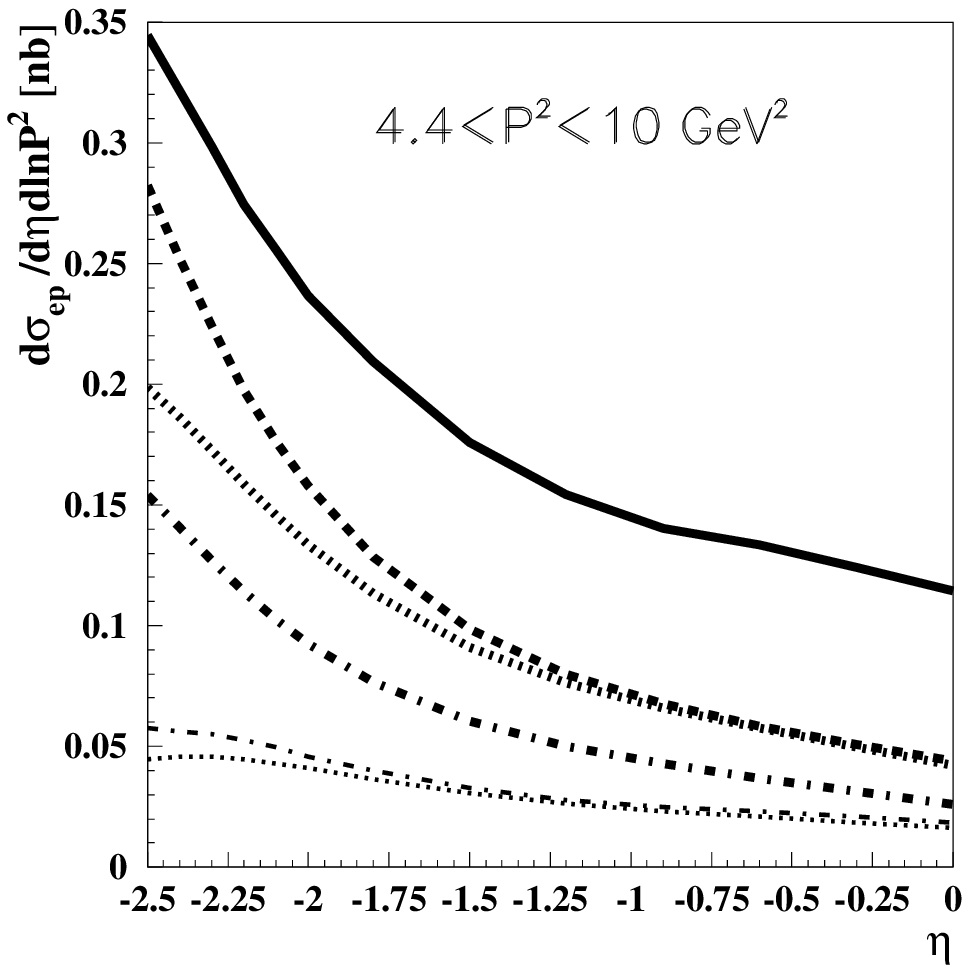,width=8.5cm,height=8.5cm}}
\put(85,0){\epsfig{file=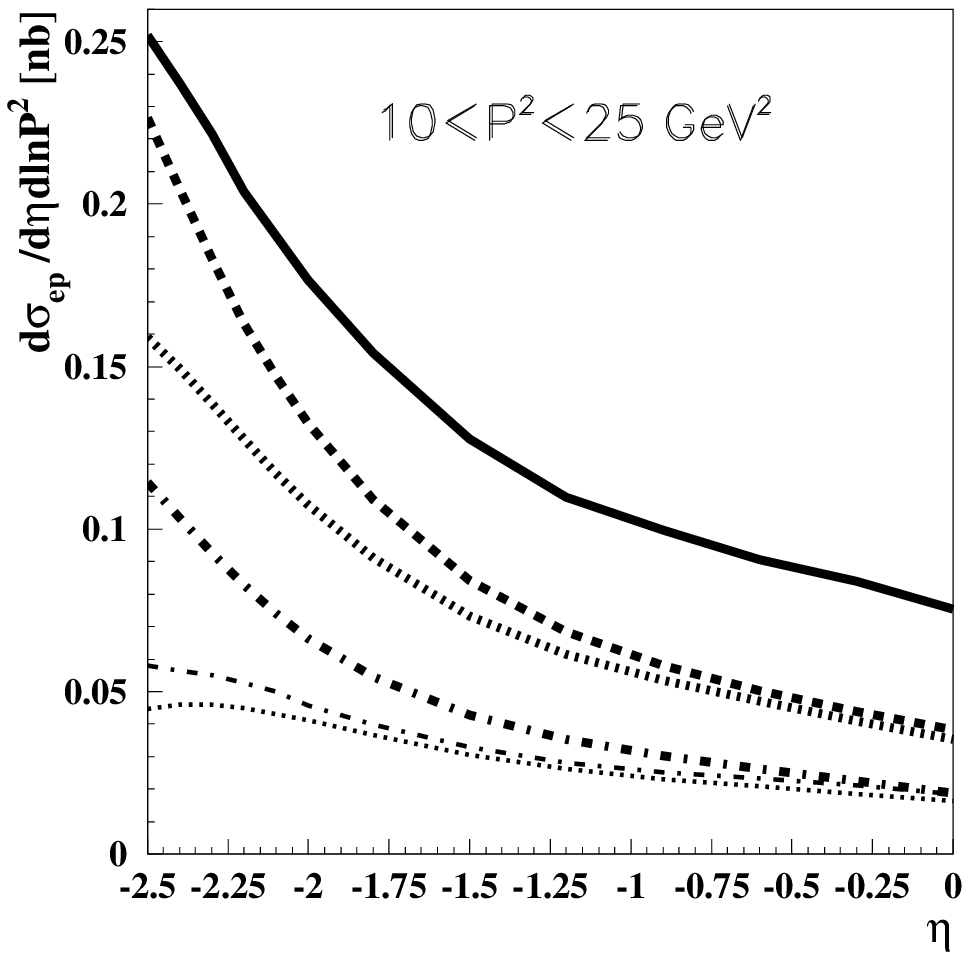,width=8.5cm,height=8.5cm}}
\end{picture}
\caption{Comparison of the complete JETVIP results taking into account
only $\gamma_T^*$ in the resolved channel with the DIR$_{\mathrm{uns}}$
ones and the convolutions of QED expressions $q^{\mathrm{QED}}_T$ and
$q^{\mathrm{QED}}_L$ with $\alpha_s^2$ partonic cross--sections. The
nontrivial effect of including $\gamma_L^*$ in the resolved channel,
given by the convolution $q_L^{\mathrm{QED}}\otimes\sigma(\alpha_s^3)$
is shown by thin dash--dotted curves.
In the case of DIR$+$RES and DIR$_{\mathrm{uns}}$ results the LO direct
contribution has been subtracted.}
\label{jetviplong1}
\end{figure}

\begin{figure}[t]\unitlength 1mm
\begin{picture}(160,120)
\put(0,60){\epsfig{file=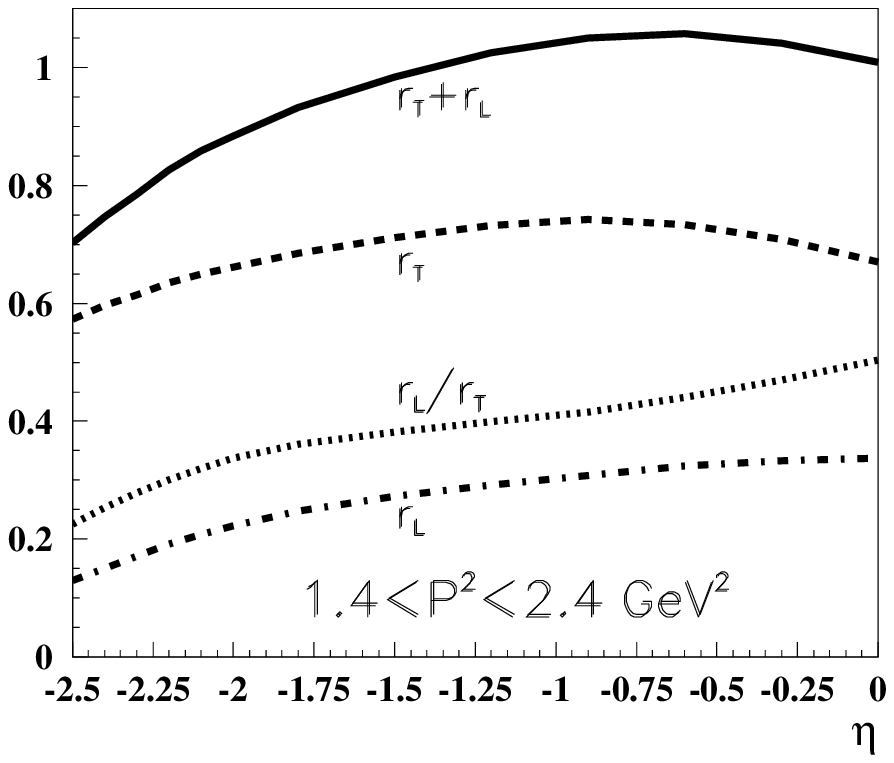,width=7.5cm}}
\put(75,60){\epsfig{file=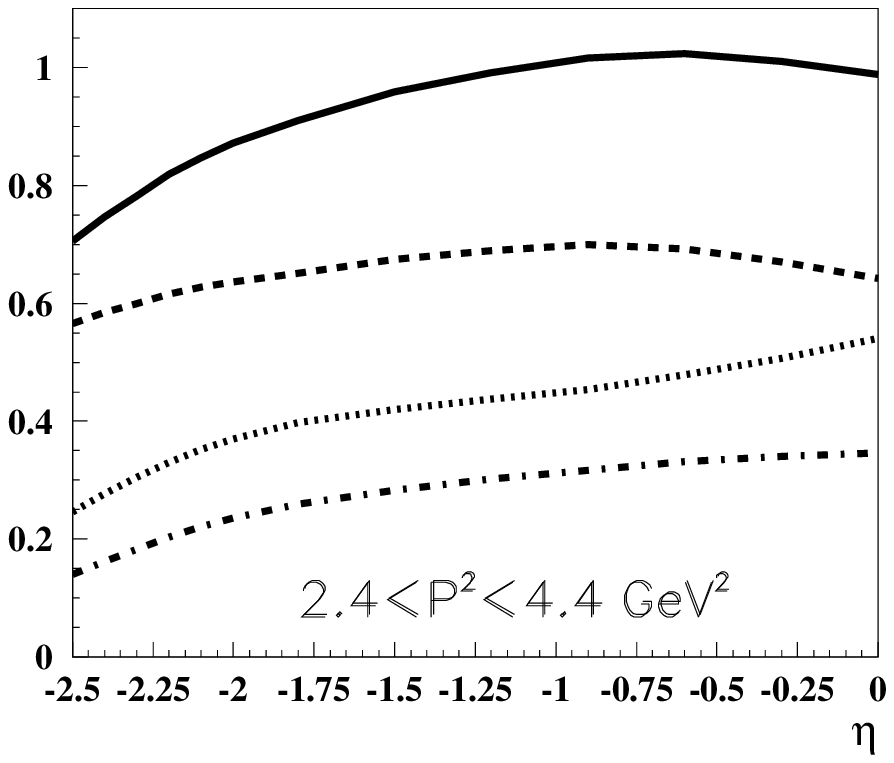,width=7.5cm}}
\put(0,0){\epsfig{file=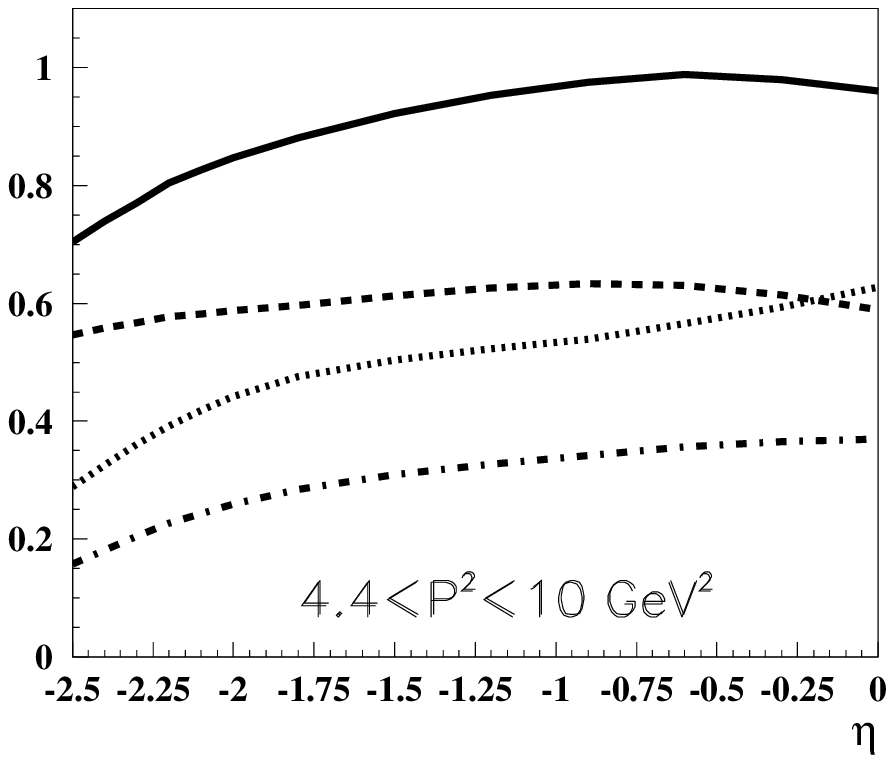,width=7.5cm}}
\put(75,0){\epsfig{file=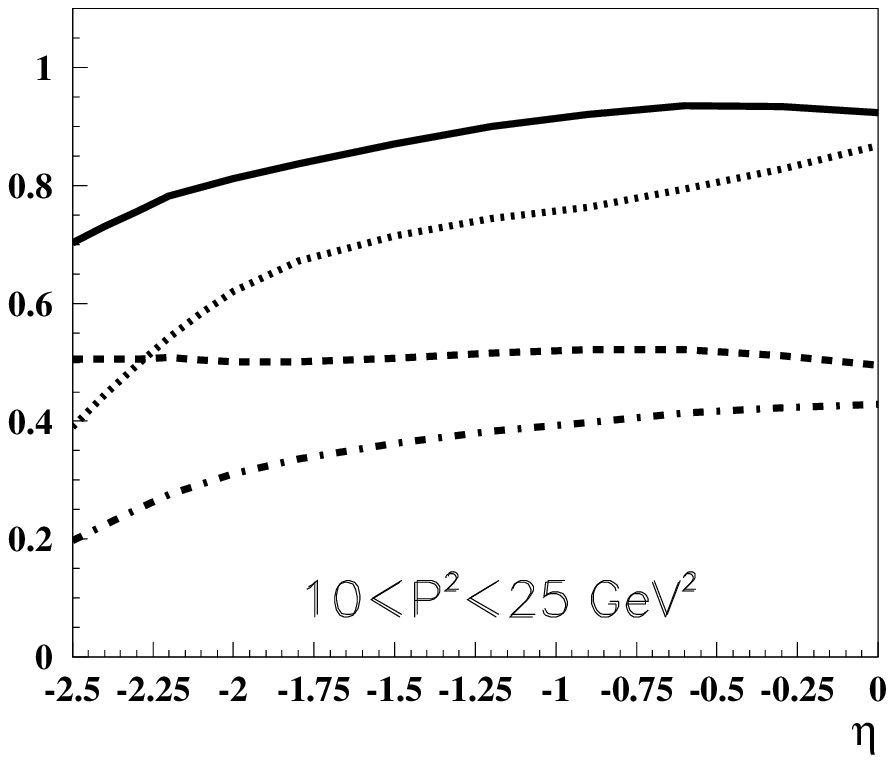,width=7.5cm}}
\end{picture}
\caption{Fractional contributions $r_L(\eta,P^2)$ and $r_T(\eta,P^2)$
together with their sum and ratio.}
\label{jetviplong3}
\end{figure}

But does it make any sense to introduce the concept of PDF of
$\gamma_L^*$? Admittedly, for interactions of virtual photons
we can stay solely within the framework of DIR$_{\mathrm{uns}}$
calculations and thus dispense with the concept of PDF of virtual
photons at all. On the other hand, as argued in
\cite{smarkem1,smarkem2,smarkem3},
the effects incorporated in the transverse part of resolved photon
component of JETVIP are numerically large.
In particular, we have emphasized the importance of including in the
resolved photon component of JETVIP the $\alpha_s^3$ partonic
cross--sections. These are not included in exact $\alpha\alpha_s^2$
DIR$_{\mathrm{uns}}$ calculations and in part of accessible
kinematical range more than double the resolved
photon contribution to dijet production at HERA compared to the
contribution of the $\alpha_s^2$ partonic cross--sections.

The same effect can be expected for $\gamma_L^*$ as the NLO
DIR$_{\mathrm{uns}}$ calculations contain at the order
$\alpha\alpha_s^2$ exact matrix elements
which include both transverse and longitudinal polarization of the
target photon. To illustrate the importance of including the
effects of $\gamma_L^*$ we compare in Fig. \ref{jetviplong1} the
convolutions $q_L^{\mathrm{QED}}\otimes\sigma(\alpha_s^2)$ and
$q_L^{\mathrm{QED}}\otimes\sigma(\alpha_s^3)$ with the
convolution $q_T^{\mathrm{QED}}\otimes\sigma(\alpha_s^2)$.
In addition, we overlay the
complete NLO DIR$+$RES and DIR$_{\mathrm{uns}}$ results, from
which, however, the LO direct photon contribution has been
subtracted. Fig. \ref{jetviplong1} shows that the
contributions of $\gamma_L^*$, though smaller, are nevertheless
comparable to those of $\gamma_T^*$, in particular for $\eta$ close
to $\eta\simeq 0$. Moreover, in the region $\eta\gtrsim -1.75$
the sum of the contributions
$(q^{\mathrm{QED}}_T+q^{\mathrm{QED}}_L)\otimes \sigma(\alpha_s^2)$
approximates remarkably well the exact
$\alpha\alpha_s^2$ DIR$_{\mathrm{uns}}$ calculations.
The excess of the exact results over this sum in the region
$\eta\lesssim -1.75$ is primarily due to the fact that the
$\alpha\alpha_s^2$ DIR$_{\mathrm{uns}}$ calculations contain beside
the tree level diagrams describing the production of three final state
partons, also one loop corrections to two parton final states, which
contribute predominantly at large negative $\eta$.

The message of Fig. \ref{jetviplong1} is quantified by plotting in
Fig. \ref{jetviplong3} the ratia
\begin{equation}
r_k(\eta,P^2)\equiv \frac{q^{\mathrm{QED}}_k\otimes
\sigma^{\mathrm{res}}(\alpha_s^2)}
{\sigma^{\mathrm{DIR}}_{\mathrm{uns}}(\alpha\alpha_s^2)},~~k=T,L
\label{ratia}
\end{equation}
of the contributions of $\gamma_T^*$ and $\gamma_L^*$, as well as
their sum, to the
$\alpha_s^2$ part of the DIR$_{\mathrm{uns}}$ results.
The ratio of the contributions of $\gamma^*_L$ and $\gamma^*_T$ is
above $1/4$ throughout the considered $\eta$ range and above $1/2$
in the region $\eta\simeq 0$. Within the DIR$_{\mathrm{uns}}$
calculations at the order $\alpha\alpha_s^2$ in the kinematical region
relevant for HERA, $\gamma_L^*$ is thus comparable in importance to
$\gamma_T^*$.
\begin{figure}\centering
\epsfig{file=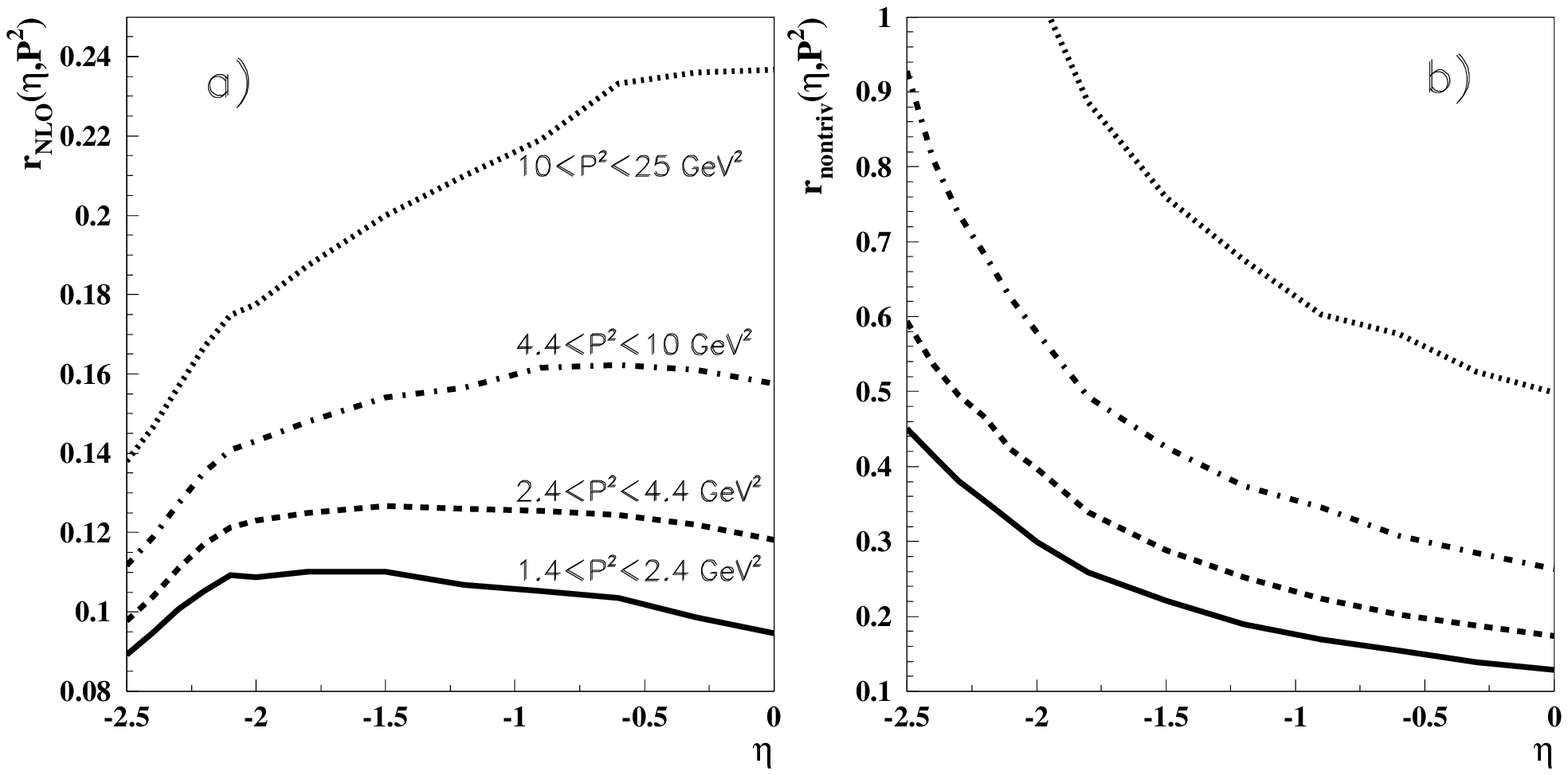,width=12cm}
\caption{The ratia $r_{\mathrm{NLO}}(\eta,P^2)$
(defined in (\ref{ratia})) and
$r_{\mathrm{nontriv}}(\eta,P^2)$ plotted as functions of $\eta$.}
\label{jetviplong}
\end{figure}

The preceding discussion illustrates the importance of the
contributions of $\gamma_L^*$, but as the $\alpha\alpha_s^2$
DIR$_{\mathrm{uns}}$ calculations include them exactly, the
genuine nontrivial effect of introducing the concept of PDF of
$\gamma_L^*$ is given in Fig. \ref{jetviplong1} by the thin
dashed-dotted curves, denoting the convolutions
$q^{\mathrm{QED}}_L\otimes\sigma(\alpha_s^3)$, which, similarly
to those of $\gamma_T^*$, are not included in
NLO DIR$_{\mathrm{uns}}$ calculations. However, as the current
version of JETVIP takes into account in the resolved channel only
the transverse virtual photons, they are not included
even in the full DIR$+$RES calculations. The net nontrivial effect
of introducing the concept of PDF of $\gamma_L^*$ into JETVIP
is then quantified by plotting in Fig. \ref{jetviplong}a the ratio
\begin{equation}
r_{\mathrm{NLO}}(\eta,P^2)
\equiv \frac{q^{\mathrm{QED}}_L\otimes
\sigma^{\mathrm{res}}(\alpha_s^3)}
{\sigma^{\mathrm{DIR}+\mathrm{RES}}}.
\label{rnlo}
\end{equation}
Also by this measure the contributions of $\gamma_L^*$ are sizable.
This net effect is much larger when the convolution
$q^{\mathrm{QED}}_L\otimes \sigma^{\mathrm{res}}(\alpha_s^3)$
is compared to the difference of DIR$+$RES and DIR$_{\mathrm{uns}}$
JETVIP results, measuring the nontrivial aspects
of the concept of PDF of $\gamma_T^*$ and corresponding to the
gap between the thick solid and dashed curves in
Fig. \ref{jetviplong1}. As shown in Fig. \ref{jetviplong}b, the
corresponding ratio, denoted $r_{\mathrm{nontriv}}(\eta,P^2)$, is
large, particularly for $\eta$ close to lower edge $\eta=-2.5$.

\section{How to measure partonic content of $\gamma_L^*$}
In principle there is no obstacle to extracting partonic content of
the virtual photon from experimental data by analyzing dijet production
at two different values of $y$. This procedure is analogous to that
involved in measuring the longitudinal structure function
$F^{\mathrm{p}}_L(x,Q^2)$ of real hadrons, which requires performing the
measurement at two different collisions energies. Although straightforward
in principle, no such direct measurement of $F_L^{\mathrm{p}}$ has been
performed at HERA, primarily for technical reasons related to changing the
proton energy. For extraction of the partonic content of $\gamma_L^*$
no such change of beam energies is necessary and it suffices to perform
the analysis of dijet cross--sections at two different values of $y$.
In practice, however, the
separation of the contributions of $\gamma_T^*$ and $\gamma_L^*$ is
not that simple, because it relies on different $y$ dependencies of
the corresponding fluxes (\ref{fluxT}-\ref{fluxL}) at large $y$.
This in turn requires measuring jet cross--sections in narrow bins
centered at two different values $y_1$ and $y_2$ instead of integrating
over the whole interval of accessible $y$, which at HERA spans
typically $0.05\lesssim y\lesssim 0.9$. Optimizing the bin width and
choice of the values $y_1,y_2$ is crucial for the success of such
extraction.

\section{Summary and conclusions}
We have demonstrated the importance of including in hard collisions
the contributions of the longitudinal polarization of the target
virtual photon. In QED these contributions
are fully calculable and their onset is determined by the ratio
$P^2/m^2$ of photon virtuality $P^2$ and fermion mass $m^2$.
The inclusion of target $\gamma_L^*$ is indispensable for good
quantitative agreement of QED predictions with existing LEP data.

In QCD gluon radiation off the quarks or antiquarks coupling to
$\gamma_L^*$ is expected to modify simple QED formulae and, in addition,
generate gluons inside $\gamma_L^*$. In this exploratory study we,
nevertheless, neglected these effects and used the purely QED formula
for quark distribution function of $\gamma_L^*$.
The numerical relevance of $\gamma_L^*$ has been illustrated within
the framework of LO analysis of observables $F_2^{\gamma}$ and
$D_{\mathrm{eff}}$ as well as within the NLO calculations of dijet
production at HERA. Better theoretical understanding of the structure
of $\gamma_L^*$ is, however, needed for more reliable evaluation
of these effects.

\vspace*{0.2cm}
\noindent
{\Large \bf Acknowledgment:}
We are grateful to J. Cvach, Ch. Friberg,
B. P\"{o}tter, I. Schienbein and A. Valk\'{a}rov\'{a}
for interesting discussions
concerning the structure and interactions of longitudinal
virtual photons. This work was supported
in part by the Grant Agency of the Academy od Sciences of the Czech
Republic under grants No. A1010821 and B1010005.

\end{document}